\documentclass[pra,twocolumn,showpacs,preprintnumbers,amsmath,amssymb,superscriptaddress,aps,10pt]{revtex4-2}
\usepackage{graphicx}
\usepackage{color}
\usepackage{amsmath}
\usepackage{subfigure}
\usepackage{epstopdf}
\usepackage{bbm}
\usepackage{multirow,array}
\usepackage{makecell}
\usepackage[mathscr]{eucal}
\newcommand{\RNum}[1]{\uppercase\expandafter{\romannumeral #1\relax}}
\usepackage[colorlinks,linkcolor=blue,citecolor=blue,urlcolor=blue,hyperindex,breaklinks]{hyperref}
\usepackage{appendix}
\usepackage{subfigure}
\usepackage{bm}

\begin{document}

\title{Quantum Speed Limit in Terms of Coherence Variations}

\author{Zi-yi Mai}
\author{Chang-shui Yu}%
 \email{ycs@dlut.edu.cn}
\affiliation{%
 School of Physics, Dalian University of Technology, Dalian
116024, P.R. China
}%

\date{\today}

\begin{abstract}
Coherence is the most fundamental quantum resource in quantum information processing. How fast a physical system gets coherence or decoherence is a critical ingredient. We present an attainable quantum speed limit based on the variation of quantum coherence subject to a dynamical process. It indicates that for a 2-dimensional quantum state, one can always find corresponding dynamics driving it to evolve along the geodesic to another state with certain coherence variation. As applications,  we study the coherence quantum speed limits of the dephasing and dissipative dynamics. It is shown that the dephasing dynamics can saturate our coherence quantum speed limit, and the decoherence of the state with identical populations will be faster than others. However, the dissipative dynamics have the opposite behavior. In addition, we illustrate a stronger tightness of our bound for the mentioned dynamics by comparison.
\end{abstract}

\maketitle

\section{introduction}
Quantum resources, including quantum coherence, entanglement, nonlocality, etc., play an essential role in quantum information processing tasks (QIPTs). In this sense, QIPTs can be understood as generating and consuming quantum resources. However, QIPTs imply that the generation of consumption of quantum resources should be as quick as possible to save QIPT time and avoid detrimental disturbance or decoherence, and so on. How fast can the resources be changed? Quantum speed limit (QSL) can answer this question by considering the least evolution time between two states with a certain amount of resources.

QSL was originally raised to describe the least time for the evolution between two states. The most typical QSL is the Mandelstam-Tam (MT) bound \cite{mandelstam1945uncertainty}
\begin{equation}\label{MT}
    \tau\geq\tau_{\mathrm{MT}}^\bot=\frac{\pi}{2\Delta E},
\end{equation}
A time-independent Hamiltonian provides a lower bound of the needed time to drive an arbitrary pure state to its orthogonal state. In other words, for the fixed energy variance, MT bound is the minimum evolution time by optimizing over all the potential state pairs and the Hamiltonian. Later, Margolus and Levitin introduced a new bound (ML bound) as \cite{MARGOLUS1998188}
\begin{equation}\label{ML}
    \tau\geq\tau_{\mathrm{ML}}^\bot=\frac{\pi}{2E},
\end{equation}
where $E$ is the expected value of the system energy. Both the MT and ML bounds are attainable for the initial state with equal weight
superposition of two eigenstates of the Hamiltonian \cite{MARGOLUS1998188,PhysRevLett.103.160502}.

{Up to now, QSL has attracted increasing interest and has been generalized in different scenarios \cite{frey2016quantum, Deffner_2017} including the QSLs for the mixed states \cite{PhysRevA.67.052109,zhang2014quantum}, time-dependent Hamiltonian \cite{PhysRevLett.65.1697,PhysRevLett.120.060409}, the geometric understandings \cite{PhysRevA.82.022107,PhysRevA.86.016101,PhysRevLett.123.180403} and so on \cite{PhysRevLett.129.140403}.  In particular, the QSL has been generalized to the open systems \cite{PhysRevA.78.012308, PhysRevLett.110.050403, PhysRevA.103.022210, Funo_2019,PhysRevA.94.052125, PhysRevA.95.022115, Campaioli2019tightrobust, PhysRevA.108.052207, PhysRevA.103.062204, PhysRevResearch.2.023299, PhysRevA.98.042132}. It is shown that  the lower bound of evolution time $\tau$ for a initial state $\rho_0$ and target state $\rho_\tau$  can be given as $\tau\geq d(\rho_0,\rho_\tau)/\langle d(\rho_t,\rho_{t+dt})/dt\rangle_\tau$, where $d$ is the distance of the two states and $\langle d(\rho_t,\rho_{t+dt})/dt\rangle_\tau=\frac{1}{\tau}\int_0^\tau d(\rho_t,\rho_{t+dt})$ is   the time-average evolution speed \cite{PhysRevX.6.021031}. For example, Refs. \cite{PhysRevLett.110.050402, PhysRevLett.111.010402} developed the geometrical QSL based on the Bures angle and explained the evolution speed as the quantum Fisher information. Ref. \cite{PhysRevA.95.052104} presented the QSL bound based on the trace distance and studied the environmental effect on the saturation of the QSL bound. The non-Markovian effect \cite{PhysRevA.91.032112, PhysRevA.93.020105, PhysRevLett.111.010402,sun2015quantum, PhysRevA.96.012105}, the effect of the Hamiltonian of an open quantum system \cite{PhysRevLett.115.210402,PhysRevA.95.052104, Lan_2022, 10.1063/5.0078207, ZHENG2024107315} and the role of coherence \cite{PhysRevA.93.052331,12,mohan2022quantum} are also addressed for the QSL. QSL has even been extended to the classical systems \cite{PhysRevLett.120.070401, PhysRevLett.120.070402}.  QSLs are also widely studied in various quantum processes such as optimal control \cite{PhysRevLett.103.240501}, quantum metrology \cite{1,2,3}, quantum battery \cite{PhysRevA.106.042436,PhysRevA.104.042209}, precision thermometry \cite{Campbell_2018}, state preparation \cite{PhysRevA.107.052608} and so on \cite{PhysRevLett.127.100404, PhysRevX.12.011038, PhysRevA.102.042606, PhysRevA.97.052333}. }
Recently, Ref. \cite{campaioli2022resource} considered the QSL of resource variation rather than the evolution between state pairs. Later, many studies were carried out for speed limit bound of the resource variation, including coherence \cite{PhysRevA.109.052443, Mohan_2022} and entanglement \cite{PhysRevA.107.022430, PhysRevA.106.042419,pandey2023fundamentalspeedlimitsentanglement, PhysRevA.107.052419}.

Quantum coherence is the most typical feature of quantum mechanics, distinguishing it from the classical world. It is also a crucial resource in many applications \cite{engel2007evidence, Plenio_2008,collini2010coherently,lloyd2011quantum,li2012witnessing,huelga2013vibrations, PhysRevLett.107.273001, PhysRevLett.113.150402,narasimhachar2015low,cwiklinski2014towards,lostaglio2015description, PhysRevX.5.021001,rebentrost2009role,witt2013stationary, PhysRevLett.115.020403, PhysRevB.84.113415, PhysRevLett.113.140401, PhysRevA.80.022324, PhysRevLett.116.160407, PhysRevLett.115.020403,yu2016total}. How fast does a physical system get coherent or decoherent? In some cases, this question potentially reveals the minimum time for the transition between quantum and classical features of a system. Although Ref. \cite{mohan2022quantum} provided a lower bound on the time required for coherence changes with von Neumann entropy as a coherence measure, whether the lower bound is attainable remains open. We will revisit the question in terms of a different coherence measure, { and we find that the fastest decoherence speed is attainable under the purely Markovian dephasing channel.}

This paper presents a QSL of the coherence variation regarding the skew information as the coherence measure \cite{PhysRevA.95.042337}. Our bound is attainable for any 2-dimensional state undergoing proper dephasing dynamics. As applications, we study the coherence of QSLs for the dephasing and dissipative dynamics. We find that the dephasing dynamics can induce faster decoherence for the states with equal populations, {and we further discuss the environmental effect on saturating our bound}, but the dissipative dynamics have the opposite behavior. We also derive the conditions for the saturations of our QSL. This paper is organized as follows. We start with a brief introduction of the coherence measure. Then, we derive a quantum speed limit bound based on the coherence measure. Then, we apply the coherence QSLs to the dephasing and dissipative dynamics and the attainability and tightness of our QSL. Finally, we summarize the paper with some discussions.

\section{coherence speed limit}
To begin with, let's first introduce the coherence measure based on the skew information \cite{PhysRevA.95.042337}.
Under the framework of the quantification of the coherence \cite{PhysRevLett.113.140401}, the coherence in terms of the skew information is defined as \cite{PhysRevA.95.042337}
\begin{equation}\label{c_measure}
    C(\rho)=\min_{\sigma\in\mathcal I}\left[1-A^2(\rho,\sigma)\right]
\end{equation}
subject to the given basis $\left\{\left\vert k \right \rangle\right\}_{k=1}^N$, where  $A(\rho,\sigma)=\mathrm{Tr}\sqrt{\rho}\sqrt{\sigma}$ is the quantum affinity \cite{Liang_2019,PhysRevA.69.032106}. It is found that incoherent states can be written as $\sigma=\sum_{k=1}^N p_k\left\vert k\right\rangle\left\langle k\right\vert$. In particular, the coherence measure in Eq. (\ref{c_measure}) can be analytically calculated for any given finite dimension.
\begin{figure}[htb]
  \centering
  \subfigure[]{\includegraphics[width=0.45\linewidth]{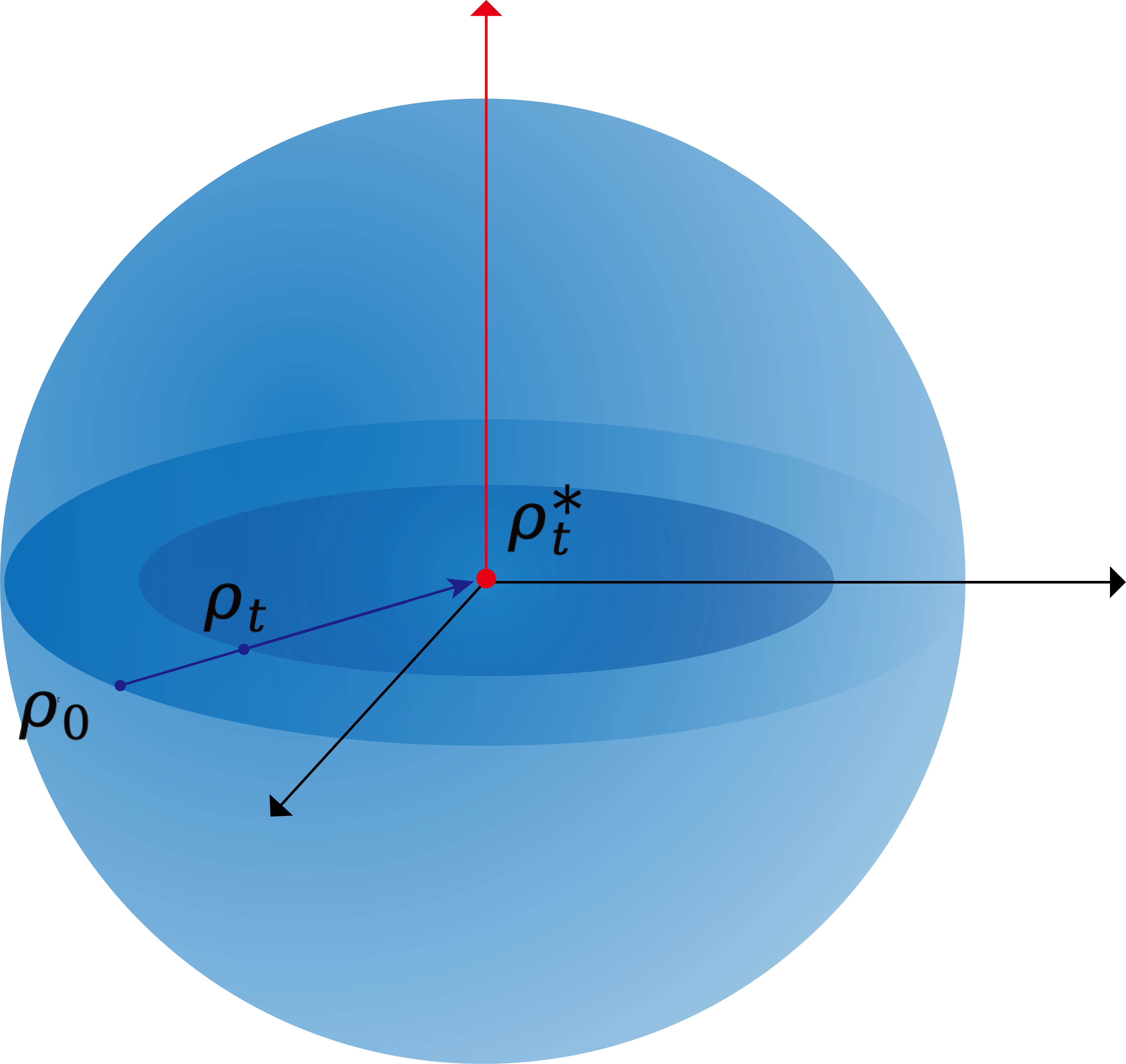}}
  \subfigure[]{\includegraphics[width=0.45\linewidth]{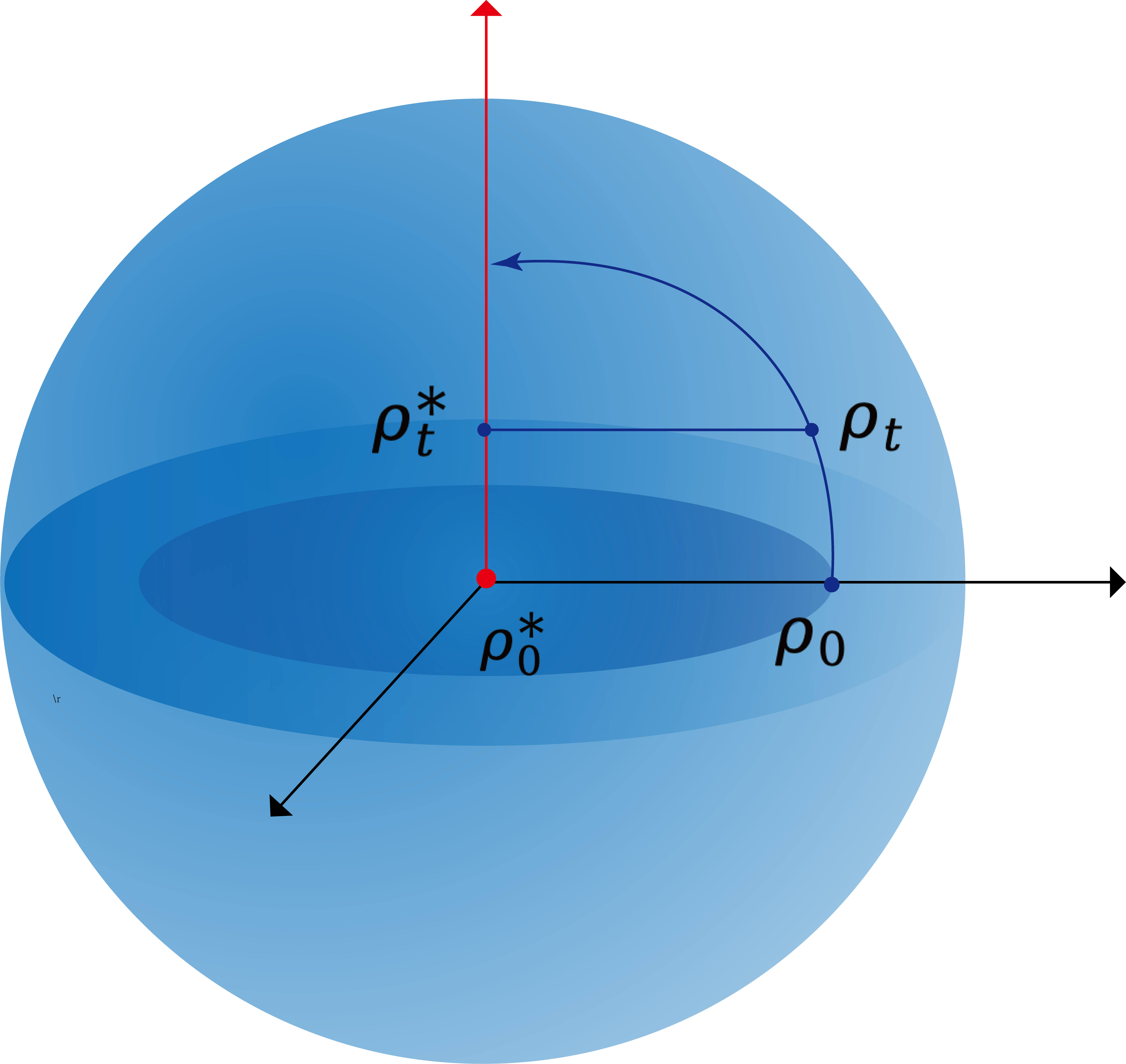}}
  \caption{The evolution trajectory in the Bloch representation. The dark blue arrow is the evolution trajectory, and the red arrow indicates the incoherent set. (a) the evolution trajectory coincides with a geodesics connecting $\rho_0$ and $\rho_0^\star$, and $\rho_t^\star=\rho_0^\star$ is satisfied for $\forall t\in[0,\tau]$. (b) evolution trajectory under a non-optimal decoherence channel.}\label{geo_fig}
\end{figure}

Based on the above-given coherence measure $C(\rho)$, we'd like to employ the metric of state distance as
\begin{equation}\label{theta}
\Theta(\rho,\sigma)=\arccos A(\rho,\sigma).
\end{equation}
It is obvious that  if $\sigma$ in  Eq. (\ref{theta}) is the closest incoherent state to $\rho$, namely, $\sigma$ is the incoherent state maximizing $\Theta(\rho,\sigma)$ in Eq. (\ref{theta}), one can easily relate the distance with the coherence measure $C(\rho)$ as
\begin{equation}
\begin{split}
    \min_{\sigma\in\mathcal I}\Theta(\rho,\sigma)&=\arccos\left\{\max_{\delta\in\mathcal I}A(\rho,\sigma)\right\}\\
    &=\arccos\sqrt{1-\min_{\sigma\in\mathcal I}\left[1-A^2(\rho,\sigma)\right]}\\
    &=\arccos\sqrt{1-C(\rho)}.
\end{split}
\end{equation}
$C(\rho)$ is the sine value of the minimum angle between $\rho$ and the incoherent states subject to Eq. (\ref{theta}). Even though $\min_{\sigma\in\mathcal I}\Theta(\rho,\sigma)$ monotonically depends on the coherence $C(\rho)$,  $\min_{\sigma\in\mathcal I}\Theta(\rho,\sigma)$ is an `addressed' coherence instead of a strict coherence measure. For convenience, we use $\rho^\star$ to represent the closest incoherent state to $\rho$.

Next, we'd like to consider the dynamic evolution of $\rho_t$ in the time interval $t\in[0,\tau]$. If $C(\rho_0)>C(\rho_\tau)$ one can obtain the following inequality:
\begin{equation}\label{ineq1}
    \begin{split}
        &\arccos\sqrt{1-C(\rho_0)}-\arccos\sqrt{1-C(\rho_\tau)}\\
        = &\Theta(\rho_0,\rho_0^\star)-\Theta(\rho_\tau,\rho_\tau^\star)\leq \Theta(\rho_0,\rho_\tau^\star)-\Theta(\rho_\tau,\rho_\tau^\star)\\
        \leq&\Theta(\rho_0,\rho_\tau)\leq\int_0^\tau dt\Theta(\rho_t,\rho_{t+dt}),
    \end{split}
\end{equation}
The final two inequalities are obtained using the triangle inequality for distance function $\Theta$. Consider the infinitesimal time interval $dt$, one can get the distance between the initial state $\rho_t$ and the final state $\rho_{t+dt}$ as
\begin{equation}\label{metric}
    \Theta^2(\rho_t,\rho_{t+dt})=\mathrm{Tr}\left(\frac{d}{dt}\sqrt{\rho_t}\right)^2,
\end{equation}
which is the Wigner-Yanase metric and is also a quantum extension to the classical Fisher information \cite{PhysRevA.69.032106}, {we present the derivation of the metric in Appendix C.} Especially for the closed system, the Wigner-Yanase metric is reduced to the skew information \cite{PhysRevA.103.022210, PhysRevX.6.021031}. For the open system, one can obtain a similar understanding of the  Wigner-Yanase metric to the closed system, as shown in Appendix A.

Similarly, for the case of $C(\rho_0)<C(\rho_\tau)$, we have
\begin{equation}\label{ineq2}
\begin{split}
        &\arccos\sqrt{1-C(\rho_\tau)}-\arccos\sqrt{1-C(\rho_0)}\\
= &\Theta(\rho_\tau,\rho_\tau^\star)-\Theta(\rho_0,\rho_0^\star)\leq\Theta(\rho_\tau,\rho_0^\star)-\Theta(\rho_0,\rho_0^\star)\\
\leq&\Theta(\rho_0,\rho_\tau)\leq\int_0^\tau dt\Theta(\rho_t,\rho_{t+dt}).
\end{split}
\end{equation}
Summarizing Eq. (\ref{ineq1}) and Eq. (\ref{ineq2}), one will directly arrive at the following theorem.

\textbf{Theorem 1}.-{\textit{For a dynamical evolution from the state $\rho_0$ to $\rho_\tau$, the time $\tau$ required for the coherence variation $\Delta_C$ is lower bounded by
\begin{equation}\label{csl}
    \tau\geq\tau_{\mathrm{CSL}}=\frac{\left\vert\Delta_C\right\vert}{\left\langle\sqrt{\mathrm{Tr}\left(\frac{d}{dt}\sqrt{\rho_t}\right)^2}\right\rangle_\tau}=\frac{\left\vert\Delta_C\right\vert}{\left\langle\sqrt{\frac{1}{4}\mathcal{I}_F+2\mathcal{I}_{W-Y}}\right\rangle_\tau},
\end{equation}
where $\Delta_C= \arccos\sqrt{1-C(\rho_\tau)}-\arccos\sqrt{1-C(\rho_0)}$  and   $\left\langle A\right\rangle_\tau=\frac{1}{\tau}\int_0^\tau dt A$ denotes the time average quantity. In particular, for the eigendecomposition $\rho_t=U_t \Lambda_t U^\dag_t$ with $\left[\Lambda_t\right]_{ij}=\lambda_j\delta_{ij}$, $\mathcal{I}_F=4\sum_j\left(\frac{d}{dt}\sqrt{\lambda_j}\right)^2$ and $\mathcal{I}_{W-Y}=-\frac{1}{2}\mathrm{Tr}\left[\sqrt{\rho_t}, H_t\right]^2$ are the classical Fisher information and the Wigner-Yanase skew information with $H_t=i\dot{U}_tU_t^\dagger$}.}

\textbf{Proof}.- The inequality in Eq. (\ref{csl}) is the direct result of combining Eq. (\ref{ineq1}) and Eq. (\ref{ineq2}). The second equality holds based on Appendix A. \hfill$\square$

Eq. (\ref{csl}) is the main result of this paper. It gives the lower bound of required time for coherence variation for any dynamical process. The minimum time among all potential dynamical processes is the QSL subject to a given coherence variation. It is mainly shown that the coherence 'variation speed' is the collective contributions of the classical Fisher information and the skew information, which quantifies the sensitivity of the state $\rho_t$ to a CPTP map due to classical and quantum effects, respectively \cite{PhysRevLett.122.010505}.
In the next section, we will show that our bound is attainable, which means a dynamical process exists, converting a state to another state with the given coherence variation in the exact evolution time $\tau_{\mathrm{CSL}}$.

Our QSL of coherence shows an intuitive geometrical picture sketched in FIG. \ref{geo_fig} in the Bloch representation. As is shown in FIG. \ref{geo_fig} (a), we draw the shortest evolution trajectory from a given state $\rho_0$ towards another state $\rho_t$ and even to an incoherent state. In this case, the speed limit inequality given in Eq. (\ref{csl}) is saturated, and the evolution trajectory coincides with the geodesics. In this evolution process, one can find that the optimal incoherent state for the coherence is always the same as the one for the initial state, i.e., for $\forall t\in[0,\tau]$, $\rho_t^\star=\rho_0^\star$. On the contrary, FIG. \ref{geo_fig} (b) demonstrates the evolution trajectory deviating from the geodesics. This evolution trajectory is not optimal for decoherence; in this case, the speed limit bound is unsaturated.

\section{Applications and the attainability}
{\textit{Dephasing dynamics.}}-  To demonstrate the attainability, let's consider a concrete example as an application. Suppose that a two-level atom interacts with a bosonic reservoir, then the Hamiltonian governing the evolution of the total system is
\begin{equation}
 H_{\mathrm{tot}}=\frac{1}{2}\omega_0\sigma_z+\sum_j\omega_jb_j^\dagger b_j+\sum_jg_j\sigma_zb_j^\dagger+\mathrm{h.c.},   \label{dephase Ham}
\end{equation}
where $\omega_0$ is the atomic transition frequency, $b_j$ is the annihilation operator of the $j$th mode in the bosonic reservoir, $\omega_j$ is the frequency of the $j$th mode harmonic oscillator, and $g_j$ is the coupling strength of atom and the $j$th mode of the reservoir. In Schr\"{o}dinger representation, one can obtain the master equation for the atomic system as \cite{breuer2002theory}
\begin{equation}\label{dephasing}
    \dot\rho_t=-i[H_0,\rho]+\frac{\gamma_t}{2}\left(\sigma_z\rho_t\sigma_z-\rho_t\right),
\end{equation}
where $H_0=\frac{1}{2}\omega_0\sigma_z$ is the free Hamiltonian, and $\gamma_t$ denotes the dephasing rate. {Note that $\gamma_t$ can take different expressions in the different approximations, which can cover the non-Markovian and Markovian cases \cite{breuer2002theory}. In the Markovian case, $\gamma_t$ typically takes a positive constant. } Let the initial state be
\begin{equation}\label{ini_state}
    \rho_0=\left(\begin{array}{ll}
        1-\rho_{11} & \rho_{01}\\
        \rho^*_{01} & \rho_{11}
    \end{array}\right),
\end{equation}
one will immediately solve Eq. (\ref{dephasing}) and obtain the state $\rho_t$ as
\begin{equation}\label{dephasing_rhot}
    \rho_t=\left(\begin{array}{ll}
        1-\rho_{11} & \rho_{01}(t)\\
        \rho_{01}^*(t) & \rho_{11}
    \end{array}\right),
\end{equation}
with {$\rho_{01}(t)=\rho_{01}e^{-\int_0^t dt'\gamma_{t'}-i\omega_0t}$.}

To show the attainable QSL, we first consider the case $\rho_{11}=\frac{1}{2}$.
In this case, one can calculate the square root of Eq. (\ref{dephasing_rhot}) and get
\begin{equation}\label{dephasing_srhot}
    \sqrt{\rho_t}=\frac{1}{2}\left(\begin{matrix}
        \sqrt{p_1}+\sqrt{p_2} & \left(\sqrt{p_1}-\sqrt{p_2}\right)e^{i\phi}\\
        \left(\sqrt{p_1}-\sqrt{p_2}\right)e^{-i\phi} & \sqrt{p_1}+\sqrt{p_2}
        \end{matrix}\right),
\end{equation}
where $p_1=\frac{1}{2}+\left\vert\rho_{01}(t)\right\vert$ and $p_2=1-p_1$ are the eigenvalues of the density matrix $\rho_t$, and $\phi=Arg\left\{\rho_{01}(t)\right\}$. Thus, according to Theorem 1, one can calculate the metric as
{
\begin{equation}\label{dephasing_vt}
\begin{split}
    \Theta^2(\rho_t,\rho_{t+dt})=\mathrm{Tr}\left(\frac{d}{dt}\sqrt{\rho_t}\right)^2=\frac{1}{4}\mathcal{I}_F+2 \mathcal{I}_{W-Y},
    \end{split}
\end{equation}
where the classical Fisher information $\mathcal{I}_F$ and the Wigner-Yanase skew information $\mathcal{I}_{W-Y}$ are given as
\begin{equation}\label{fisher}
  \mathcal{I}_F=4\sum_{j=0}^1\left(\frac{d}{dt}\sqrt{p_j}\right)^2=4\frac{\left\vert\rho_{01}(t)\right\vert^2}{{1-4\left\vert\rho_{01}(t)\right\vert^2}}\gamma_t^2
\end{equation}
and
\begin{equation}\label{wy}
  \mathcal{I}_{W-Y}=-\frac{1}{2}\mathrm{Tr}\left[\sqrt{\rho_t},H_0\right]^2
  =\frac{\omega_0^2}{2}\left(\frac{1}{2}-\sqrt{\frac{1}{4}-\left\vert\rho_{01}(t)\right\vert^2}\right),
\end{equation}
respectively. It is obvious that the Wigner-Yanase skew information $\mathcal{I}_{W-Y}$ given in Eq. (\ref{wy}) corresponds to the contribution of the free Hamiltonian $H_0$. The classical Fisher information $\mathcal{I}_F$ given in Eq. (\ref{fisher})  denotes the contribution of the dephasing process. }

{Integrating the square root of Eq. (\ref{dephasing_vt}) from $0$ to $\tau$, one can immediately find that
\begin{eqnarray}
  &&\int_0^\tau dt\sqrt{\frac{1}{4}\mathcal{I}_F+2 \mathcal{I}_{W-Y}}\geq\int_0^\tau dt\sqrt{\frac{1}{4}\mathcal{I}_{F}}\notag\\
  &&\geq \left\vert\int_0^\tau dt\frac{\vert\rho_{01}(t)\vert}{\sqrt{1-4\vert\rho_{01}(t)\vert}}\gamma_t\right\vert=\left\vert\int_0^\tau dt\frac{\frac{d}{dt}\vert\rho_{01}(t)\vert}{\sqrt{1-4\vert\rho_{01}(t)\vert^2}}\right\vert\label{18}\\
  &&=\left\vert\frac{1}{2}\left(\arcsin 2\left\vert\rho_{01}(\tau)\right\vert-\arcsin 2\left\vert\rho_{01}\right\vert\right)\right\vert\notag\\
&&=\left\vert\arcsin\sqrt{\frac{1}{2}-\sqrt{\frac{1}{4}-\left\vert \rho_{01}(\tau)\right\vert^2}}-\arcsin\sqrt{\frac{1}{2}-\sqrt{\frac{1}{4}-\left\vert \rho_{01}\right\vert^2}}\right\vert\notag\\
&&=\left\vert\arccos\sqrt{\frac{1}{2}-\sqrt{\frac{1}{4}-\left\vert \rho_{01}(\tau)\right\vert^2}}-\arccos\sqrt{\frac{1}{2}-\sqrt{\frac{1}{4}-\left\vert \rho_{01}\right\vert^2}}\right\vert\notag\\
 && =\left\vert\Delta_C\right\vert,\label{dephasing_dissvt}
\end{eqnarray}
where we have used the equality $\arcsin\sqrt{\frac{1}{2}-\sqrt{\frac{1}{4}-x^2}}=\frac{1}{2}\arcsin 2x$. Besides, according to Ref. \cite{PhysRevA.95.042337}, we have $C(\rho)=1-\sum_k\left\langle k\right\vert\sqrt{\rho}\left\vert k\right\rangle^2=1-\frac{1}{2}\left(\sqrt{p_1}+\sqrt{p_2}\right)^2
=\frac{1}{2}-\sqrt{\frac{1}{4}-\left\vert\rho_{01}(t)\right\vert^2}$, which corresponds to the final result in  Eq. (\ref{dephasing_dissvt}). The first inequality saturates for the vanishing W-Y skew information, which can be achieved if the two energy levels are degenerate, i.e., $\omega_0=0$. Of course, one can also, at least in principle, consider a particular engineered noise $\xi(t)$ similar to Ref. \cite{PhysRevA.95.052104} to eliminate the frequency $\omega_0$ and mimic a purely dephasing dynamics. }

\begin{figure}[htbp]
\includegraphics[width=0.48\linewidth]{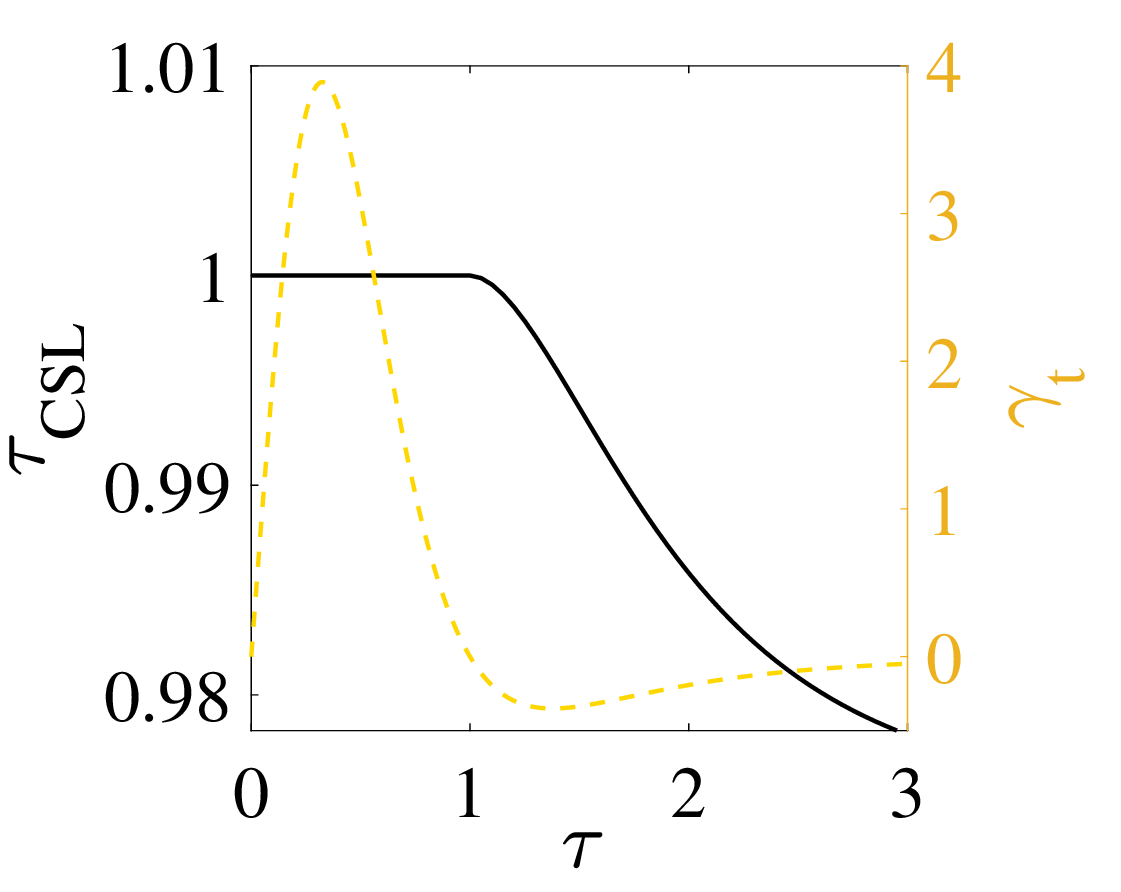}
\includegraphics[width=0.49\linewidth]{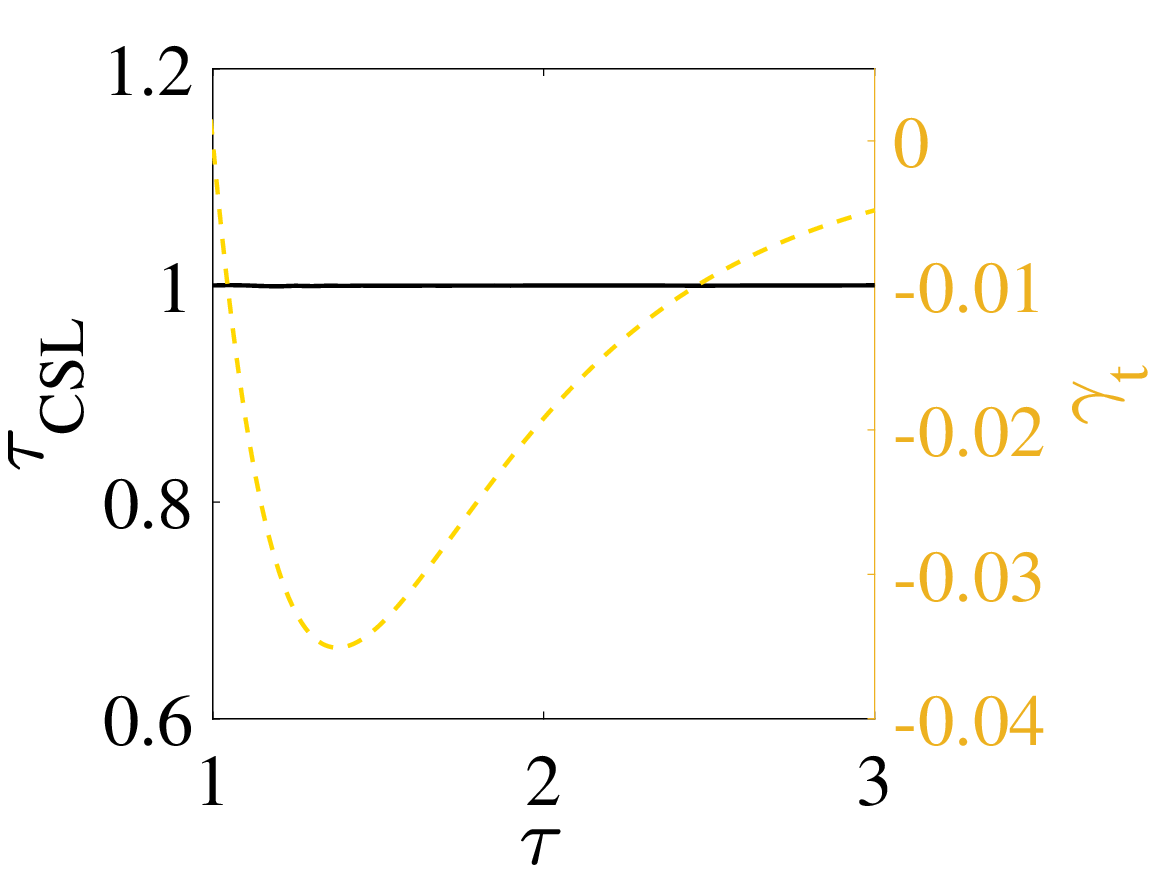}
\caption{(Left) The ratio $\tau_{\mathrm{CSL}}/\tau$ and dephasing rate $\gamma_t$ $vs$ evolution time $\tau$ . (Right) The ratio $\tau_{\mathrm{CSL}}/\tau$ and dephasing rate $\gamma_t$ $vs$ evolution time $\tau$, and let the initial instant locates at $t^\star=1$. Both are considered in the non-Markovian process ($k=4$).}
\label{non_markov}
\end{figure}

{The second inequality of Eq. (\ref{18}) saturates if $\gamma_t$ doesn't change its sign in the internal $t\in[0,\tau]$. Note that $\gamma_t$ can take negative values within some time intervals due to the information exchange between the system and environment in a non-Markovian regime. Thus, Eq. (\ref{dephasing_dissvt}) can safely saturate in the Markovian regime. However, some non-Markovian decoherence processes can also saturate Eq. (\ref{dephasing_dissvt}). Let's consider the dephasing rate \cite{breuer2002theory,PhysRevA.104.042202}
\begin{equation}
  \gamma_t=\int d\omega J(\omega)\coth\left(\frac{\omega}{2k_BT}\right)\frac{\sin\omega t}{\omega},
\end{equation}
where $J(\omega)$ denotes the Ohmic-like spectral density of the reservoir, and in the low-temperature limit, $J(\omega)$ can be given as
\begin{equation}
  J(\omega)=\frac{\omega^k}{\omega_c^{k-1}}e^{-\frac{\omega}{\omega_c}}
\end{equation}
with cutoff energy $\omega_c$, and $k$ describing the sub-Ohmic ($k<1$), Ohmic ($k=1$) and super-Ohmic ($k>1$) environment. Under the zero temperature limit, the dephasing rate can be expressed as
\begin{equation}
\gamma_t=\omega_c\left(1+\omega_c^2t^2\right)^{-k/2}\Gamma\left(k\right)\sin\left[k\arctan\left(\omega_c t\right)\right]
\end{equation}
with the Gamma function $\Gamma$. We provide the numerical result of $\tau_{\mathrm{CSL}}/\tau$ with different evolution times $\tau$ for the super-Ohmic environment under the non-Markovian effect in FIG. \ref{non_markov}, from which one can see that our bound saturates within the time interval ($\tau\leq 1$) with the positive dephasing rate but unsaturated when it turns to a negative value as increasing evolution time ($\tau>1$). The above cases mainly belong to the decoherence. The coherence-generating process can also happen along the geodesics at some time intervals. In FIG. \ref{non_markov}, one can find that the dephasing rate becomes negative from the moment $t^\star=1$, and our QSL bound is saturated. To sum up, one can find that the non-Markovian effect (with $\gamma_t$ changing sign in $[0,\tau]$) and the free Hamiltonian $H_0$ can increase the length of the evolution trajectory and hence make the dynamics deviate the geodesics.
}

The above calculation indicates that the $2\times 2$ density matrix with identical diagonal entries is the optimal state to be accelerated to the maximum decoherence speed. $\rho_t$ approaches the incoherent state set along the geodesics under purely dephasing dynamics {with the Markovian regime and the vanishing free Hamiltonian}, meanwhile the geodesics connecting $\rho_0$ and $\rho_\tau$ travels through $\rho_0^\star$, i.e., $\rho_t^\star=\rho_0^\star$, $\forall t\in[0,\tau]$.

We want to emphasize that for any $2\times 2$ density matrix, there always exists an equal-population representation, where the diagonal entries of the density matrix are the same, i.e., $\rho_{11}=\frac{1}{2}$ in Eq. (\ref{ini_state}). Such a representation can be easily realized by applying the unitary transformation
$U=\frac{1}{\sqrt{2}}\left(\begin{array}{cc}
        1 & ie^{i\phi}\\
        1 & -ie^{i\phi}\\
\end{array}\right).$
We can impose the above dephasing process {under the Markovian regime} in this representation and obtain similar results. In this sense, one can conclude that for any initial state, we can always find proper dephasing dynamics such that the system's coherence is degraded along the geodesics of the state evolution. Namely, our coherence QSL is attainable.

\begin{figure}
    \centering
    \includegraphics[width=0.49\linewidth]{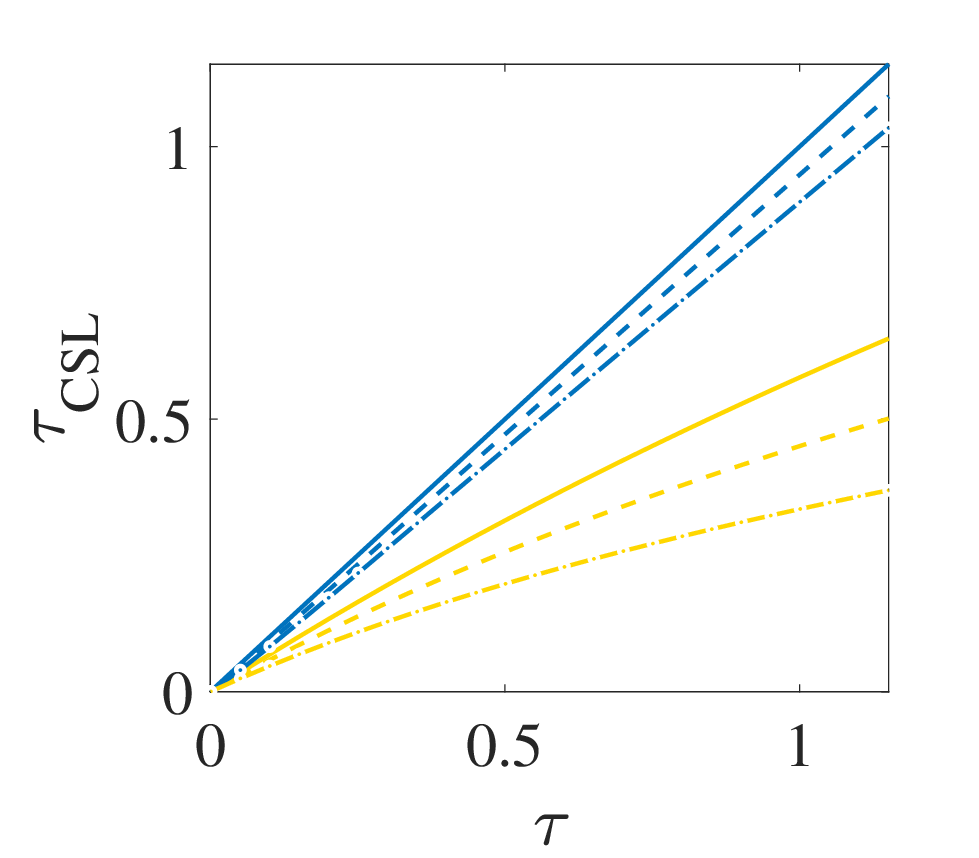}
        \includegraphics[width=0.49\linewidth]{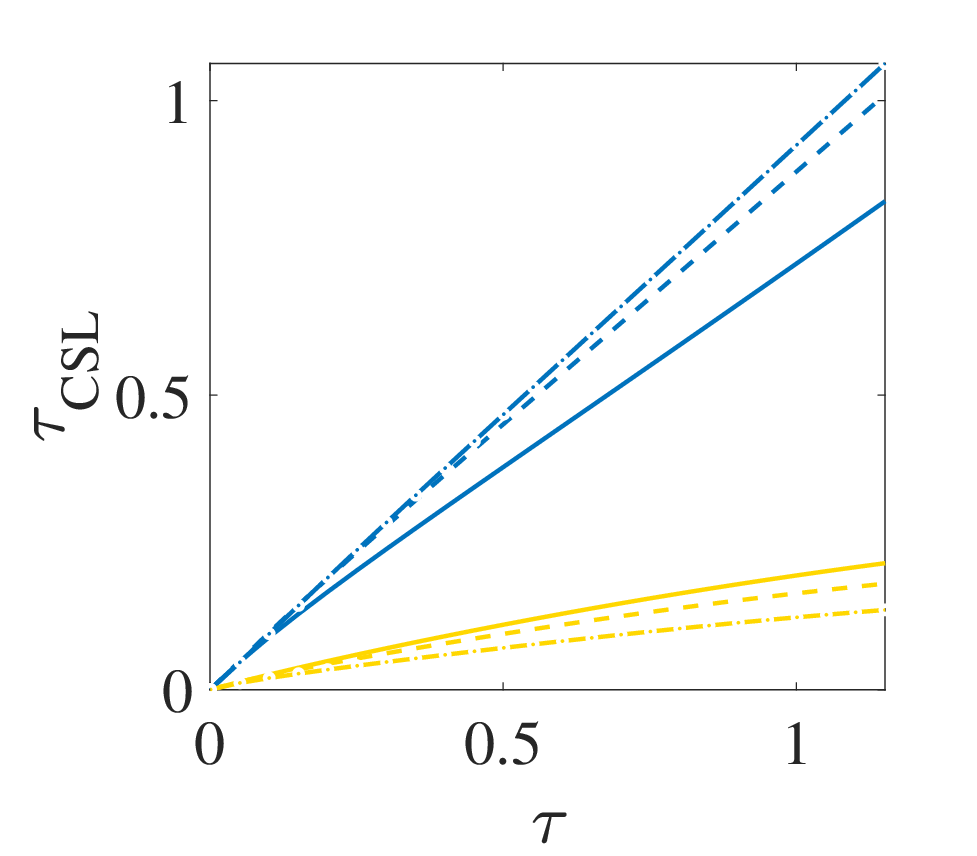}
    \caption{The actual evolution time $\tau$ under the purely dephasing channel (Left) and the amplitude damping channel (Right) $vs$ the speed limit time $\tau_{\mathrm{CSL}}$. The blue lines and the yellow lines are obtained by our coherence QSL time $\tau_{\mathrm{CSL}}$  and that in Ref. \cite{mohan2022quantum}, respectively, and the solid, dash and dash-dot lines represent the parameters $\theta=\frac{\pi}{2}, \frac{\pi}{3}$ and $\frac{\pi}{4}$ of the initial state, respectively.}
    \label{dephasing_fig1}
\end{figure}

We also numerically study the dephasing dynamics with different initial states. We set $\rho_{11}=\sin^2\frac{\theta}{2}$ and $\rho_{01}=\sin\frac{\theta}{2}\cos\frac{\theta}{2}$. The decay rate is selected as $\gamma_t=2$, {and the frequency difference is zero $\omega_0=0$}. FIG. \ref{dephasing_fig1} (Left) shows the result, and meanwhile, we also attach the numerical result presented in Ref. \cite{mohan2022quantum} for comparison. It can be seen that our speed limit bound shows preferable tightness for this model. The dephasing dynamics subject to the states with non-identical diagonal entries show a relatively slow decoherence speed. Namely, they always lead to $\tau>\tau_{\mathrm{CSL}}$,  { which means these evolution trajectories deviate the geodesics, and the optimal decoherence dynamics allow the initial state evolves along the geodesics towards its closest coherent state (See FIG. \ref{geo_fig}.)} This result implies that the non-identical population can distinctly influence decoherence speed under dephasing dynamics {which is attributed to the population-dependent coherence measure.}

{\textit{Dissipative dynamics}}.- To better understand the coherence of QSL, we further consider the dissipative dynamics. Suppose a two-level atom interacts with a leaky single-mode cavity \cite{breuer2002theory, PhysRevA.55.2290, PhysRevLett.111.010402}.  The Hamiltonian of this system reads
\begin{equation}
H_{\mathrm{tot}}=\frac{1}{2}\omega_0\sigma_z+\sum_j\omega_jb_j^\dagger b_j+\sum_j g_j\sigma_+b_j+\mathrm{h.c.},\end{equation}
where various operators are defined analogous to those in Eq. (\ref {dephase Ham}). If the environment is a vacuum at the initial moment, one can derive
the master equation of the reduced system as
\begin{equation}
    \dot\rho_t=\frac{\gamma_t}{2}\left(2\sigma_-\rho_t\sigma_+-\left\{\sigma_+\sigma_-,\rho_t\right\}\right),\label{equa}
\end{equation}
where $\rho_t$ is the density matrix of the system of interest, and  $\gamma_t$ is the time-dependent decay rate. If the initial state of the system is given by Eq. (\ref{ini_state}), then one can solve Eq. (\ref{equa}) and get the density matrix $\rho_t$ as
\begin{equation}
    \rho_t=\left(\begin{matrix}
        q_{11}(t) & q_{01}(t)\\
        q_{01}^*(t) & 1-q_{11}(t)
    \end{matrix}\right)
\end{equation}
with $q_{11}(t)=1-\rho_{11}e^{-\int_0^t dt'\gamma_{t'}}$ and $q_{01}(t)=\rho_{10}e^{-\int_0^t dt'\gamma_{t'}/2}$. Here we study $\tau_{\mathrm{CSL}}$ dependent on the evolution time $\tau$ numerically. The figure is shown in FIG. \ref{dephasing_fig1} (Right), where the initial state and decay rate parameters are the same as the numerical example in the dephasing model. Our bound is larger than the bound in Ref. \cite{mohan2022quantum}, which indicates our bound has better tightness. Besides, according to the numerical result in FIG. \ref{dephasing_fig1} (Right), the initial state with identical entries ($\theta=\pi/2$) shows lower decoherence speed compared with the cases of $\theta=\pi/3$ and $\theta=\pi/4$.  This result is exactly opposite to the results in the dephasing dynamics. The initial states with the same diagonal entries show the maximum decoherence speed for the dephasing case. However, in the JC model, the initial state with asymmetric diagonal entries exhibits a large decoherence speed. This phenomenon can be understood as follows. For the initial states with small non-diagonal entries, the main contribution to the evolution speed comes from the variation of the diagonal entries. The evolution trajectory derived from mere variation of the diagonal entries of the density matrix is closer to the geodesic, so even if the initial states with asymmetric diagonal elements do not decohere along the most direct path, they exhibit greater dynamical speed.
We compare the bound presented by Ref. \cite{mohan2022quantum}.

\textit{{Attainability}.}- To gain further insight into the attainability of our coherence speed limit bound, we consider the geodesics $\rho_t$, $t\in[0,\tau]$ such that $\rho_t^\star=\rho_0^\star$ for $\forall t$, the geodesics has been explicitly given in Refs. \cite{jenvcova2004geodesic,gibilisco2003wigner}. It is shown that the geodesics connecting $\rho_0$ and $\rho_\tau$ reads
\begin{equation}
    \rho_t=\frac{\left[(1-p_t)\sqrt{\rho_0}+p_t\sqrt{\rho_\tau}\right]^2}{\mathrm{Tr}\left[(1-p_t)\sqrt{\rho_0}+p_t\sqrt{\rho_\tau}\right]^2},
\end{equation}
the uniqueness of $\sqrt{\rho_t}$ indicates that
\begin{equation}\label{geod}
    \sqrt{\rho_t}=\frac{(1-p_t)\sqrt{\rho_0}+p_t\sqrt{\rho_\tau}}{\sqrt{(1-p_t)^2+p_t^2+2p_t(1-p_t)\mathrm{Tr}\sqrt{\rho_0}\sqrt{\rho_\tau}}}
\end{equation}
with monotonic real function $p_t$ satisfying $p_0=0$ and $p_\tau=1$. The space consisting of all the density matrices equipping the Wigner-Yanase metric is partly the Euclidean sphere \cite{gibilisco2003wigner}, hence the geodesics Eq. (\ref{geod}) is a linear combination of $\sqrt{\rho_0}$ and $\sqrt{\rho_\tau}$, with the normalized numerator guaranteeing the length of $\sqrt{\rho_t}$ is unit, i.e., $\mathrm{Tr}\rho_t=1$, which is an immediately consequence of the sphere geometry \cite{gibilisco2003wigner}. For integrity, we provide an alternative method to showing the geodesics connecting $\rho_0$ and $\rho_\tau$ in Appendix B, apart from Refs. \cite{jenvcova2004geodesic,gibilisco2003wigner}. The closest incoherent state to $\sqrt{\rho_t}$ is \cite{PhysRevA.95.042337}
\begin{equation}\label{28}
\begin{split}
    &\rho_t^\star=\sum_{i=1}^N\frac{\left\langle i\right\vert\sqrt{\rho_t}\left\vert i\right\rangle^2}{\sum_{j=1}^N\left\langle j\right\vert\sqrt{\rho_t}\left\vert j\right\rangle^2}\left\vert i\right\rangle\left\langle i\right\vert=\\
    &\sum_{i=1}^N\frac{\left(\left\langle i\right\vert \sqrt{\rho_0}\left\vert i\right\rangle-p_t\left\langle i\right\vert \left(\sqrt{\rho_0}-\sqrt{\rho_\tau}\right)\left\vert i\right\rangle\right)^2}{\sum_{j=1}^{N}\left(\left\langle j\right\vert \sqrt{\rho_0}\left\vert j\right\rangle-p_t\left\langle j\right\vert \left(\sqrt{\rho_0}-\sqrt{\rho_\tau}\right)\left\vert j\right\rangle\right)^2}\left\vert i\right\rangle\left\langle i\right\vert.
    \end{split}
\end{equation}
From Eq. (\ref{ineq1}) and Eq. (\ref{ineq2}), one can immediately find that the saturated CSL needs $\rho_0^\star=\rho_\tau^\star$ .  Namely, $\rho_t^\star$ (Eq. (\ref{28})) shoudn't depend on time.  Obviously, 
if 
\begin{equation}\label{attain_1}
    \left\langle i\right\vert \sqrt{\rho_t}\left\vert i\right\rangle=\left\langle j\right\vert \sqrt{\rho_t}\left\vert j\right\rangle
\end{equation}
or
\begin{equation}\label{attain_2}
    \left\langle i\right\vert \sqrt{\rho_0}\left\vert i\right\rangle=\left\langle i\right\vert \sqrt{\rho_\tau}\left\vert i\right\rangle
\end{equation} for $\forall t\in[0,\tau]$, $\forall i,j$, 
Eq. (\ref{28}) will be time-independent.

Eq. (\ref{attain_1}) shows that the matrix $\sqrt{\rho_t}$ should have the same diagonal density entries, which indicates that $\rho_t$ in 2 dimension should have identical diagonal entries, i.e., $\frac{1}{2}$, this characteristic can be only generated by the purely dephasing channel. However, this could differ for high-dimensional density matrices, which need further study due to their complexity. The density matrix $\rho_t$ will have to own some particular form to guarantee the identical diagonal entries of $\sqrt{\rho_t}$. Similarly,  Eq. (\ref{attain_2}) indicates that the diagonal entries of $\sqrt{\rho_t}$ shouldn't depend on time $t$. For a 2-dimensional density matrix, Eq. (\ref{attain_2}) means $\rho_0=\rho_\tau$, which is a trivial case.  Under the condition Eq. (\ref{attain_1}),  dephasing dynamics is the only solution to saturate our speed limit bound in 2 dimension.

\section{Conclusion and discussion}
This paper establishes a quantum speed limit for coherence variation based on the skew information coherence measure. This QSL provides a lower bound on how long it takes for a physical system to generate or lose certain coherence. The results indicate that the classical Fisher information and the Wigner-Yanase skew information contribute two important fractions to the average evaluation speed. The non-vanishing Wigner-Yanase skew information corresponding to some unitary evolution often prevents the dynamics from evolving along the geodesics, which can be further verified in Appendix D, where we have also shown that the unitary evolution cannot be along the geodesics. Hence, the free Hamiltonian (unitary evolution) slow the evolution during the coherence variation in the case we raised. In this sense, the potential candidate could be the dynamics between degenerated energy levels or some particularly designed case, as shown in the above section. In addition, we also derive the saturation condition of our QSL bound. We find one condition is that the square root of quantum states during the evolution keep identical diagonal entries. In particular, in two-dimensional systems, this saturation condition needs the state to have the same populations during the evolution. That is, our QSL bound is attainable only in a two-level system with identical populations undergoing a purely dephasing channel with the constant-sign dephasing rate. Thus, one conclusion is that coherence-generating dynamics cannot evolve along the geodesics for a two-dimensional system, but decoherence can. Roughly speaking, coherence generation should be slower than decoherence. What's more, one can find that the coherence QSL bound depends on the populations, which is intuitively attributed to the fact that the skew information coherence measure depends on the populations and simultaneously induces a population-dependent metric, but essentially, the evolution trajectory of density matrices depends on initial states as well as their populations. The dissipative dynamics can further understand this behavior. Unlike dephasing dynamics, we find that the states with the same diagonal entries exhibit less decoherence speed for the dissipative dynamics because the initial states with asymmetric diagonal entries own the evolution trajectory close to the geodesics. These two opposite results indicate the factors of decoherence for the different dynamics. Comparing our bound with that in Ref. \cite{mohan2022quantum}, our bound exhibits relatively preferable compactness. Since quantum coherence is a fundamental feature of a quantum system, our coherence QSL bound reveals the lower bound of the transition time from quantum to classical features.

\section{acknowledgements}
This work was supported by the National Natural Science Foundation of China under Grants Nos. 12175029, 12011530014.

\section*{Appendix A: THE PHYSICS OF
WIGNER-YANASE METRIC FOR OPEN
SYSTEM}
In this appendix, we show the physical meaning of the
W-Y metric. First, f r arbitrarily given time-dependent
density matrix $\rho_t$ can be decomposed as \cite{PhysRevX.12.011038}
\begin{equation}
    \rho_t=U_t\Lambda_t U_t^\dagger=\left(U_t\sqrt{\Lambda_t}U_t^\dagger\right)^2,
\end{equation}
where $U_t$ is a time-dependent unitary matrix, and $\Lambda_t=\sum_i\lambda_i(t)\vert i\rangle\langle i\vert$ with $\vert i\rangle$ denoting the $i-$th eigenvector of $\rho_0$. The uniqueness of $\sqrt{\rho_t}$ indicates that
\begin{equation}
\sqrt{\rho_t}=U_t\sqrt{\Lambda_t}U_t^\dagger=\sum_i\sqrt{\lambda_i(t)}U_t\vert i\rangle\langle i\vert U_t^\dagger,
\end{equation}
and its derivative is expressed as
\begin{equation}
    \frac{d}{dt}\sqrt{\rho_t}=\dot{U}_tU_t^\dagger\sqrt{\rho_t}+\sqrt{\rho_t}U_t\dot{U}_t^\dagger+\sum_i\left(\frac{d}{dt}\sqrt{\lambda_i(t)}\right)\vert i(t)\rangle\langle i(t)\vert,
\end{equation}
where $\vert i(t)\rangle=U_t\vert i\rangle$ is the $i-$th eigenvector of
$\rho_t$. By defining the effective Hamiltonian $H_t=i\dot{U}_tU_t^\dagger$, it's not difficult to find that $H_t$ is hermitian, then one will immediately find that
\begin{equation}
    \frac{d}{dt}\sqrt{\rho_t}=-i[H_t,\sqrt{\rho_t}]+\sum_i\left(\frac{d}{dt}\sqrt{\lambda_i(t)}\right)\vert i(t)\rangle\langle i(t)\vert,
\end{equation}
then the W-Y metric is
\begin{equation}\label{appa_metric}
\begin{split}
  \mathrm{Tr}\left(\frac{d}{dt}\sqrt{\rho_t}\right)^2=&\sum_j\left(\frac{d}{dt}\sqrt{\lambda_j(t)}\right)^2-\mathrm{Tr}\left[\sqrt{\rho_t},H_t\right]^2\\
  &+2i\mathrm{Tr}\left(\sum_j\left(\frac{d}{dt}\sqrt{\lambda_j(t)}\right)\left\vert j\right\rangle\left\langle j\right\vert\left[\sqrt{\rho_t},H_t\right]\right),
\end{split}
\end{equation}
where the final term in the second line vanishes due to the commutative $\sqrt{\rho_t}$ and $\sum_j\frac{d}{dt}\sqrt{\lambda_j(t)}\left\vert j\right\rangle\left\langle j\right\vert$. Hence
\begin{equation}\label{appa_metric1}
  \mathrm{Tr}\left(\frac{d}{dt}\sqrt{\rho_t}\right)^2=\frac{1}{4}\mathcal{I}_F+2\mathcal{I}_{W-Y},
\end{equation}
where $\mathcal{I}_F=4\sum_j\left(\frac{d}{dt}\sqrt{\lambda_j(t)}\right)^2$ and $\mathcal{I}_{W-Y}=-\frac{1}{2}\mathrm{Tr}\left[\sqrt{\rho_t}, H_t\right]^2$ are the classical Fisher information and the Wigner-Yanase skew information, respectively. The classical Fisher information and the skew information of the effective Hamiltonian quantifies the sensitivity of the state $\rho_t$ to a CPTP map due to classical and quantum effects, respectively \cite{PhysRevLett.122.010505, PhysRevX.12.011038}.

\section*{Appendix B: the geodesics concerning the Wigner-Yanase metric}
We want to refer to Refs. (\cite{jenvcova2004geodesic,gibilisco2003wigner}) to give the geodesic. Let's fi st consider a real vector space $\mathbb R^N$, $\bm{r_0}$ and $\bm{r_\tau}$ are arbitrarily given pair of normalized vectors, i.e., $\vert\bm{r_{0,\tau}}\vert=1$, and $\bm{r_0}\neq -\bm{r_\tau}$. Then
\begin{equation}
    \bm{r_t}=\frac{p_t\bm{r_0}+(1-p_t)\bm{r_\tau}}{\left\vert p_t\bm{r_0}+(1-p_t)\bm{r_\tau}\right\vert}
\end{equation}
is the normalized vector within the plane expanded by $\bm{r_0}$ and $\bm{r_\tau}$, where $p_t$ is the monotonic function satisfying $p_0=1$ and $p_\tau=0$. Obviously, on a two-dimensional plane, the following triangle inequality is always saturated:
\begin{equation}
\arccos\left\langle\bm{r_0},\bm{r_\tau}\right\rangle=\arccos\left\langle\bm{r_0},\bm{r_t}\right\rangle+\arccos\left\langle\bm{r_t},\bm{r_\tau}\right\rangle,
\end{equation}
where
\begin{equation}
\begin{split}
    \langle \bm{r_0},\bm{r_t}\rangle&=\frac{p_t+(1-p_t)\langle\bm{r_0},\bm{r_\tau}\rangle}{\sqrt{p_t^2+(1-p_t)^2+2p_t(1-p_t)\langle\bm{r_0},\bm{r_\tau}\rangle}}\equiv f_{p_t}\left(\langle\bm{r_0},\bm{r_\tau}\rangle\right)\\
    \langle \bm{r_\tau},\bm{r_t}\rangle&=\frac{p_t\langle\bm{r_0},\bm{r_\tau}\rangle+(1-p_t)}{\sqrt{p_t^2+(1-p_t)^2+2p_t(1-p_t)\langle\bm{r_0},\bm{r_\tau}\rangle}}\equiv g_{p_t}\left(\langle\bm{r_0},\bm{r_\tau}\rangle\right).
    \end{split}
\end{equation}
One should notice that the above equations are satisfied for $\forall\langle \bm{r_0},\bm{r_\tau}\rangle\in(-1,1)$ and $p_t\in[0,1]$. Hence
\begin{equation}
\arccos x=\arccos f_{p_t}(x)+\arccos g_{p_t}(x)
\end{equation}
hold for any input $x\in(-1,1)$ and $p_t\in[0,1]$. Let $x=\mathrm{Tr}\sqrt{\rho_0}\sqrt{\rho_\tau}$, it's not difficult to find that it equals to
\begin{equation}
\Theta(\rho_0,\rho_\tau)=\Theta(\rho_0,\rho_t)+\Theta(\rho_t,\rho_\tau)
\end{equation}
for $\forall p_t\in[0,1]$, it indicates that Eq. (\ref{geod}) is the geodesics connecting $\sqrt{\rho_0}$ and $\sqrt{\rho_\tau}$.

{
\section*{Appendix C: The derivation of the Wigner-Yanase metric}
To provide a clear theoretical framework, we'd like to present a derivation of the form of the Wigner-Yanase metric. First, considering an affinity function
\begin{equation}\label{appc_f}
  A(\rho_t;s)=A(\rho_t,\rho_{s})=\mathrm{Tr}\sqrt{\rho_t}\sqrt{\rho_{s}}
\end{equation}
with respect to one parameter $s\in[0,\tau]$ and $\rho_t$ is a curve $t:0\rightarrow \tau$ in the state space. When $s$ is close to $t$, one can expand $A$ to the second order with respect to $s$ as \begin{equation}\label{appc_2A}
  A(\rho_t;s)=A(\rho_t,t)+\frac{\partial}{\partial s}A(\rho_t;s)\Big\vert_{s=t}dt+\frac{1}{2}\frac{\partial^2}{\partial s^2}A(\rho_t;s)\Big\vert_{s=t}dt^2
\end{equation}
with $dt=s-t$. It can be calculated that $A(\rho_t;t)=\mathrm{Tr}\rho_t=1$, $\frac{\partial}{\partial s}A(\rho_t;s)\Big\vert_{s=t}=\mathrm{Tr}\sqrt{\rho_t}\frac{d}{dt}{\sqrt{\rho_t}}=0$ and $\frac{\partial^2}{\partial s^2}A(\rho_0;s)\Big\vert_{s=t}=\mathrm{Tr}\sqrt{\rho_t}\frac{d^2}{dt^2}\sqrt{\rho_t}=-\mathrm{Tr}\left(\frac{d}{dt}{\sqrt{\rho_t}}\right)^2$, hence
\begin{equation}\label{appc_A0}
  A(\rho_t;s)=\mathrm{Tr}\sqrt{\rho_t}\sqrt{\rho_s}=1-\frac{1}{2}\mathrm{Tr}\left(\frac{d}{dt}{\sqrt{\rho_t}}\right)^2
\end{equation}
holds for $s$ approaching $t$.  Hence for any pair of neighboring states $\rho_t$ and $\rho_{t+dt}$, the affinity is given as
\begin{equation}\label{appc_at}
\begin{split}
  A(\rho_t,\rho_{t+dt})&=1-\frac{1}{2}\mathrm{Tr}\left(\frac{d}{dt}{\sqrt{\rho_t}}\right)^2\\
  &=\cos\Theta(\rho_t,\rho_{t+dt})=1-\frac{1}{2}\Theta^2(\rho_t,\rho_{t+dt}).
  \end{split}
\end{equation}
Then, one can immediately observe that the Wigner-Yanase metric is
\begin{equation}\label{appc_metric}
  \Theta^2(\rho_t,\rho_{t+dt})=\mathrm{Tr}\left(\frac{d}{dt}{\sqrt{\rho_t}}\right)^2.
\end{equation}
}

{
\section*{Appendix D: Unitary evolution deviates the geodesics}
Let's first assume that the unitary evolution in Eq. (\ref{geod}) can be expressed as $\sqrt{\rho_t}=U_t\sqrt{\rho_0}U_t^\dagger$.  It can shown that this expression is invalid. For our assumption, it's not difficult to observe that $\mathrm{Tr}\sqrt{\rho_t}$ is invariant for $\forall t\in[0,\tau]$.  Then one can immediately get
\begin{equation}
  \mathrm{Tr}\sqrt{\rho_t}=\frac{1}{\sqrt{1-2p_t\left(1-p_t\right)\left(1-\mathrm{Tr}\sqrt{\rho_0}\sqrt{\rho_\tau}\right)}}.
\end{equation}
It can be seen that, time-independence of the term $\mathrm{Tr}\sqrt{\rho_t}$ requires $\mathrm{Tr}\sqrt{\rho_0}\sqrt{\rho_\tau}=1\Leftrightarrow\rho_0=\rho_\tau$.  Hence $\sqrt{\rho_t}=U_t\sqrt{\rho_0}U_t^\dagger$ is valid only for $U_t=\mathbb I$, which is a trivial case.
}

\bibliography{QSLcoherence_bib}
\end{document}